\documentclass[preprint,12pt]{article}

\usepackage{amssymb}

\usepackage{color}
\usepackage{threeparttable}
\usepackage[dvipdfmx]{graphicx}

\newcommand*{\red}[1]{\textcolor{black}{#1}}

\title{A two-phase model of collective memory decay with a dynamical switching point}
\author{Naoki Igarashi \and Yukihiko Okada \and Hiroki Sayama \and Yukie Sano\thanks{University of Tsukuba, Japan. sano@sk.tsukuba.ac.jp}}

\begin{document}

\maketitle

\begin{abstract}
Public memories of significant events shared within societies and groups have been conceptualized and studied as ``collective memory'' since the 1920s. 
Thanks to the recent advancement in digitization of public-domain knowledge and online user behaviors, collective memory has now become a subject of rigorous quantitative investigation using large-scale empirical data. Earlier studies, however, typically considered only one dynamical process applied to data obtained in just one specific event category. Here we propose a two-phase mathematical model of collective memory decay that combines exponential and power-law phases, which represent fast (linear) and slow (nonlinear) decay dynamics, respectively. We applied the proposed model to the Wikipedia page view data for articles on significant events in five categories: earthquakes, deaths of notable persons, aviation accidents, mass murder incidents, and terrorist attacks. 
Results showed that the proposed two-phase model compared favorably with other existing models of collective memory decay in most of the event categories. The estimated model parameters were found to be similar across all the event categories. 
The proposed model also allowed for detection of a dynamical switching point when the dominant decay dynamics exhibit a phase shift from exponential to power-law. Such decay phase shifts typically occurred about 10 to 11 days after the peak in all of the five event categories.
\end{abstract}

\section{Introduction}
Memories of past significant events, such as natural disasters and wars, that are shared by members of social groups like countries and families are called ``collective memory". 
This concept was proposed in 1925 by a sociologist Halbawchs~\cite{halbwachs1992collective}.
\red{Later, Assman classified} collective memory into communicative memory and cultural memory depending on how the memory is passed down to future generations \cite{assmann1995collective}. 
\red{Communicative memory is maintained by everyday communications such as conversations with close people.
By contrast, cultural memory is maintained by cultural formation (texts, rites, monuments) and institutional communication (recitation, practice, observance) ~\cite{assmann1995collective, candia2019universal}.}

\red{Although these were initially treated as sociological concepts,} recently, it has begun to attract attention as a target for empirical research. 
\red{How much people remember World War II is investigated by country \cite{roediger2019competing} and age \cite{Zaromb2014-he} through self-reported surveys. 
Roediger et al. examined how much people forgotten the U.S. presidents through interviews with students \cite{roediger2014forgetting} and found two different functions that characterize forgetting.}

Thanks to the recent advancement in digitization of public-domain knowledge and online user behaviors, there has been a growing effort to study collective memory quantitatively using large-scale empirical data. 
For example, Michel et al.~\cite{michel2011quantitative} investigated collective memory using the word frequencies in digitized books. Au Yeung et al.~\cite{au2011studying} measured the extent to which collective memories were retained in different countries using large data sets of news articles.
Page views and edit histories of Wikipedia articles about significant events, such as natural and man-made disasters, aviation accidents, and terrorist attacks, have been frequently used as indicators of collective memory~\cite{ferron2014beyond,kanhabua2014triggers,garcia2017memory,ferron2011collective}.
Some studies used Wikipedia not only to measure the level of collective memory but also to understand the collective nature of people in more general sense, such as revealing the relationship between page views and turnout in elections~\cite{yasseri2016wikipedia}, building a model of people's browsing behavior considering external factors~\cite{ratkiewicz2010characterizing}, and predicting the popularity of movies from page views~\cite{mestyan2013early}.
Singer et al.~\cite{singer2017we} showed that mass media and current events (30\% and 13\% of respondents, respectively) dominated the motivation for people to access Wikipedia pages. It was found that Wikipedia page view activity strongly correlates with Google search activity\cite{5591292,10.1145/2786451.2786495}. These earlier studies warrant the use of Wikipedia page views as a quantitative metric of the general user behavior on the Internet.

\red{Mathematical models were proposed to describe collective memory decay and validated with various empirical data, including Wikipedia page views.}
Candia et al.~revealed the universal nature of decay patterns on a yearly time scale~\cite{candia2019universal}.
They showed that the decay of collective memory can be modeled by a biexponential function \red{$C_{1}\mathrm{e}^{-\alpha t}+C_{2} \mathrm{e}^{-\beta t}$} using the number of citations of papers and patents as well as online attention to songs, movies, and biographies \red{of Wikipedia views on a yearly scale.}

\red{Collective memory decay have also been investigated daily time scale. 
Kim et al.~\cite{kim2021stretched} proposed a stretched exponential function $\mathrm{e}^{-(t/\alpha)^{\beta x^{-\gamma}}}$ to describe daily page views of online academic articles. They successfully depict the dynamics that decay quickly in the beginning and slowly in the latter using the stretched exponential function.}
West et al.~\cite{west2021postmortem} studied the daily collective user behavior of Twitter and news sites on the news about the deaths of celebrities between 2009 and 2014. They showed that the mention frequency can be modeled by a shifted power-law function \red{$C_{1} t^{-\alpha}+C_{2}$ with their exponents are $\alpha=1.34$ and $\alpha = 1.54$ for news and Twitter, respectively}. 
Garc\'{i}a-Gavilanes et al.~\cite{garcia2016dynamics} analyzed daily Wikipedia page view dynamics on articles of aviation accidents and found that the collective memory decays exponentially after it reaches a maximal value. \red{They proposed a segmented model that assumes separated regimes behind the dynamics.}

The earlier studies on daily collective memory decay dynamics typically considered only one dynamical process applied to data obtained in just one specific event category. Whereas a universal model~\cite{candia2019universal} was proposed for annual collective memory decay using the data for multiple event categories, such year-by-year dynamics are only relevant at a slow, historical time scale, which would not be applicable to day-to-day dynamics. There is hence a need for a universal model of collective memory decay for a faster, daily time scale.

Here we propose a \red{two-phase} decay model for collective memory of various types of significant events and evaluate its validity using Wikipedia page view data.
We compare the proposed decay model to several other existing decay models developed using data from Wikipedia~\cite{kobayashi2021modeling}, blogs~\cite{sano2013empirical}, Twitter~\cite{asur2011trends,lorenz2019accelerating}, YouTube~\cite{crane2008robust}, news sites~\cite{wu2007novelty,dezso2006dynamics}, book sales~\cite{sornette2004endogenous}, and the number of articles read~\cite{kim2021stretched}. These earlier studies modeled collective memory decay in various mathematical forms, such as power-law, exponential, and stretched exponential, to which the proposed model is compared for performance evaluation.

\section{Data and methodology}
\subsection{Data}

\begin{threeparttable}
\caption{Summary of Wikipedia page data.}
      \begin{tabular}{ccc}
        \hline
        Category  & URL of Summary Article\tnote{*}  & \# of Events  \\ \hline
        Earthquakes & \small{List\_of\_earthquakes\_2011-2020} & 74 \\
        Notable deaths & \small{Deaths\_in\_[January-December]\_[2015-2020]} & 22319 \\
        Aviation accidents & \footnotesize{Category:Aviation\_Accidents\_and\_incidents\_in\_[2015-2020]}  & 107 \\
        Mass murders & \small{Category:Mass\_murder\_in\_[2015-2020]} & 121 \\
        Terrorist attacks & \small{List\_of\_terrorist\_incidents\_in\_[2015-2020]} & 403 \\
        \hline
    \end{tabular}
    \label{Tab1} 
    
\begin{tablenotes}
\item[*] URL following: https://en.wikipedia.org/wiki/ 
\end{tablenotes}
\end{threeparttable}

In this study we analyzed collective memory decay using English Wikipedia page view data.
We selected the following five categories of significant events for analysis: earthquakes, deaths of notable persons, aviation accidents, mass murder incidents, and terrorist attacks. These events were also used in previous collective memory studies~\cite{kanhabua2014triggers,garcia2017memory,garcia2016dynamics,west2021postmortem, watanabe2020}.
For the events in these categories, the date and location of the event are precise, which allows for the collection of unambiguous time series data. We obtained the Wikipedia pages of events listed in the summary article of each category in the English version of Wikipedia. The target period of event occurrence is from July 1st, 2015, to June 30th, 2020. Table~\ref{Tab1} shows a summary of the dataset we obtained from Wikipedia.

\begin{figure}
    \includegraphics[scale=0.4]{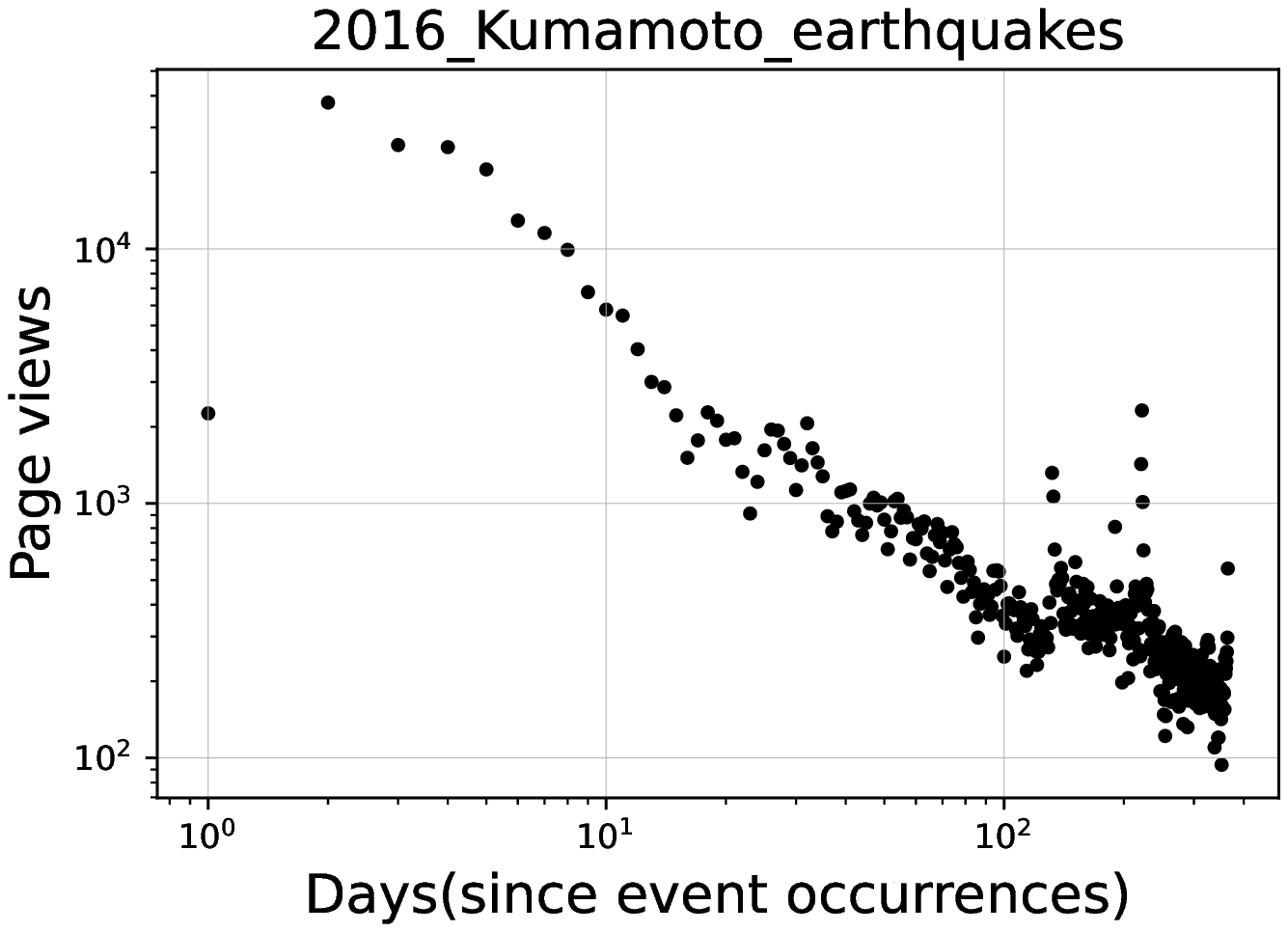}
    \includegraphics[scale=0.4]{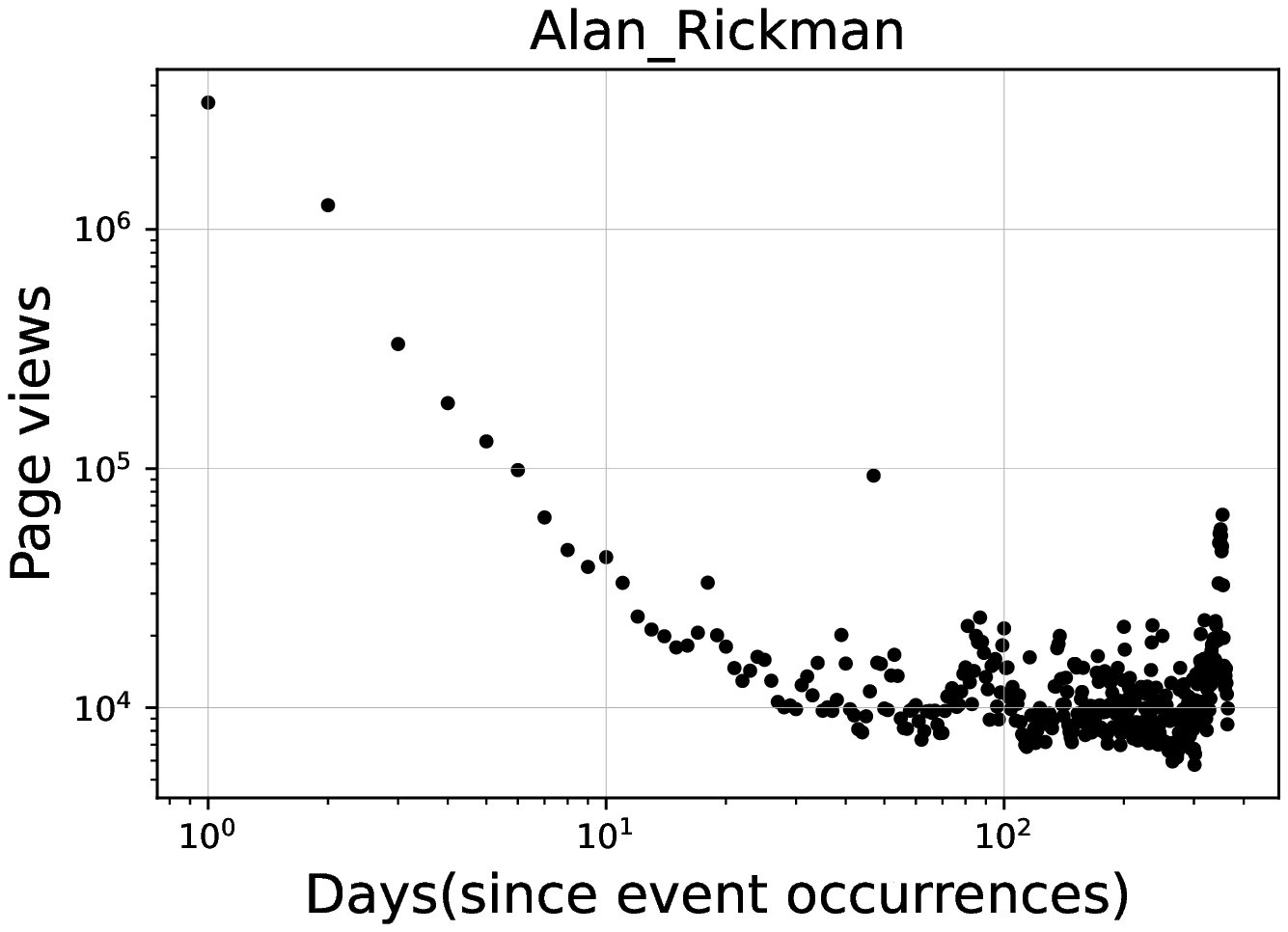}
\caption{Examples of Wikipedia page view decay.}
    \label{Fig1}  
\end{figure}

Figure~\ref{Fig1} shows two examples of Wikipedia page view decay from the event occurrence date (one for the 2016 earthquakes in Kumamoto, Japan, and the other for the death of Alan Rickman).
The two examples commonly show that the Wikipedia page views peaked around the date of the event and gradually decayed over time. In addition, the peak height of the page views (i.e., how much attention an event receives) and the decay rate (i.e., how quickly people forget it) varied greatly from event to event. 

For each of the collected Wikipedia pages, we obtained daily page view counts since the event occurrence date for 300 days from the infobox in each event Wikipedia page by using Wikimedia REST API (https://wikimedia.org/api/restv1/). The length of the data collection period was set to 300 days, shorter than one year, in order to avoid a possible ``anniversary'' page view increase toward the end of the 365-day cycle
(such increase was seen in Fig.~\ref{Fig1} right). 
If the page view peak was less than 1,000 or occurred 5 or more days after the event date, the data was excluded from the analysis since we considered such events did not trigger significant collective memory responses. With these criteria, we acquired valid page view data for 34 earthquakes, 8,684 deaths of notable persons, 43 aviation accidents, 37 mass murder incidents, and 123 terrorist attacks.

\subsection{Model}
\begin{figure}
\includegraphics[scale=0.5,angle=0]{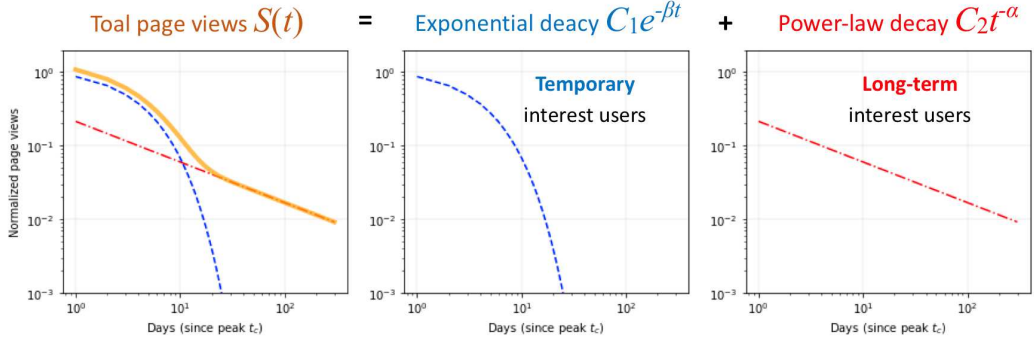}
\caption{The proposed model of collective memory decay.}
\label{Fig2}
\end{figure}

In this study, we propose a unique two-phase mathematical model of collective memory decay that combines exponential and power-law phases. 
\red{Our model does not assume a regime shift in the decay of collective memory but rather a change within the population that forms collective memory.}
First, we define the normalized daily page views $t$ days after the peak $t_c$ for each event as $S(t)={S^{\mathit{raw}}(t)}/{S^{\mathit{raw}}(0)}$, where $S^{\mathit{raw}}(t)$ is the raw number of daily page views $t$ days after the peak $t_c$ (and therefore $S^{\mathit{raw}}(0)$ is the number of page views at the peak $t_c$).
Next, we assume that there are two types of users: the first type is ``temporary interest users", whose page views decay rapidly as an exponential function of time with no interaction, and the second type is ``long interest users'', whose page views decay following a power-law function of time which implies non-trivial interactions among those users. 
Combining these two types of users determines the total number of page views in our model (Fig.~\ref{Fig2}). This model can capture the shift from ``fast decay'' to ``slow decay". The model formula is mathematically expressed as follows:

\begin{equation}
 S(t) =C_{1}\mathrm{e}^{-\beta t} + C_{2}t^{-\alpha}
\end{equation}

$C_{1}$ and $C_{2}$ are constant parameters representing the amplitudes of the two decay dynamics. $\beta$ is the decay rate of the initial exponential decay, and $\alpha$ is the decay rate of the mid- to long-term power-law decay. This proposed model is different from the models of the previous research, and the idea that the basic properties of the user can be divided into two distinct groups is also unique to our research. 

To evaluate the validity of this model, we quantitatively compared the accuracy of the proposed model with that of other models in the previous research, including bi-exponential~\cite{kim2021stretched}, stretched exponential~\cite{candia2019universal}, and shifted power-law~\cite{west2021postmortem} by measuring the coefficient of \red{determination} $R^{2}$ and the Akaike Information Criterion ($AIC$).

\section{Results}
\subsection{Model fitting}

\begin{figure}
    \includegraphics[keepaspectratio,scale=0.35]{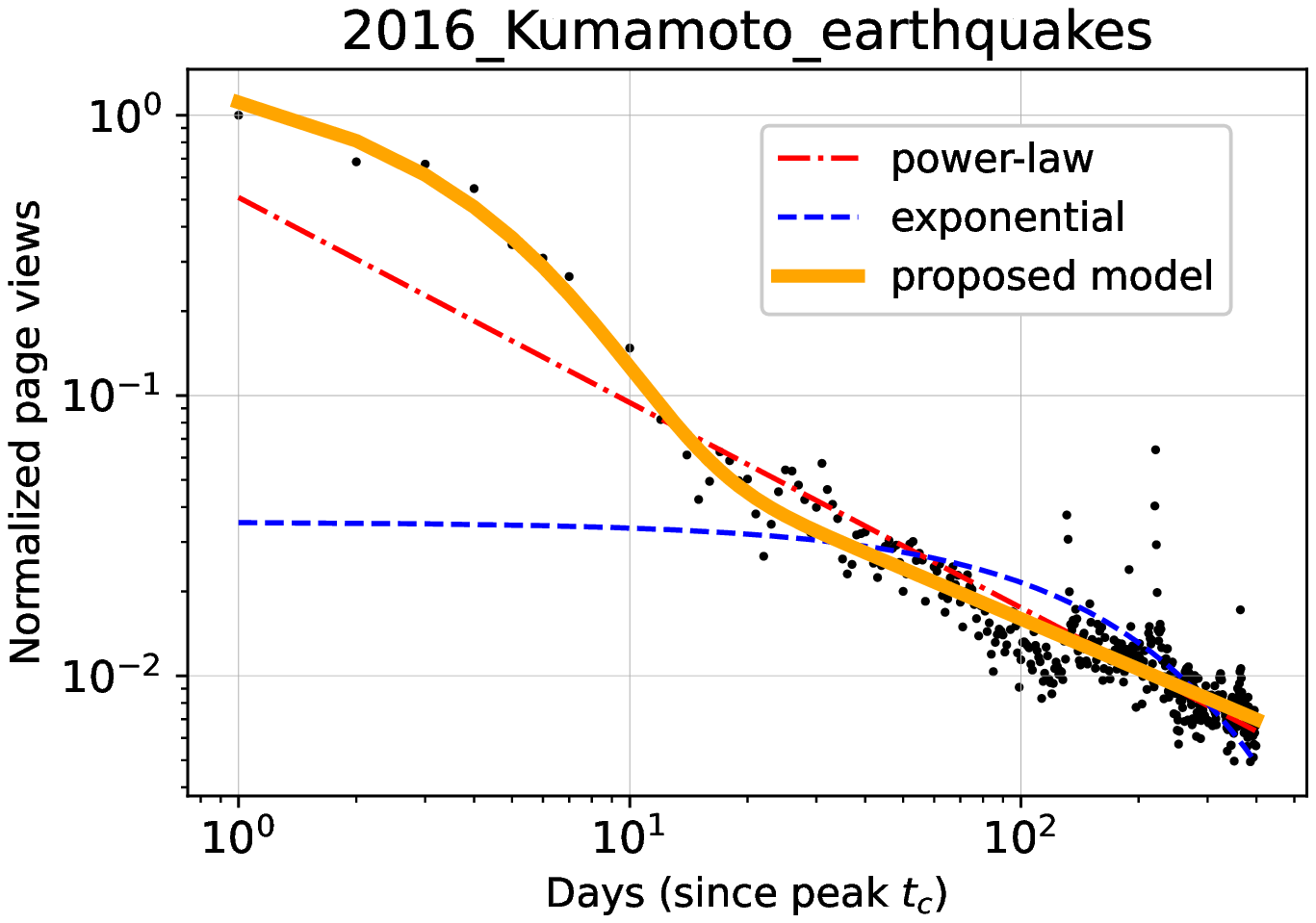}
    \includegraphics[keepaspectratio,scale=0.35]{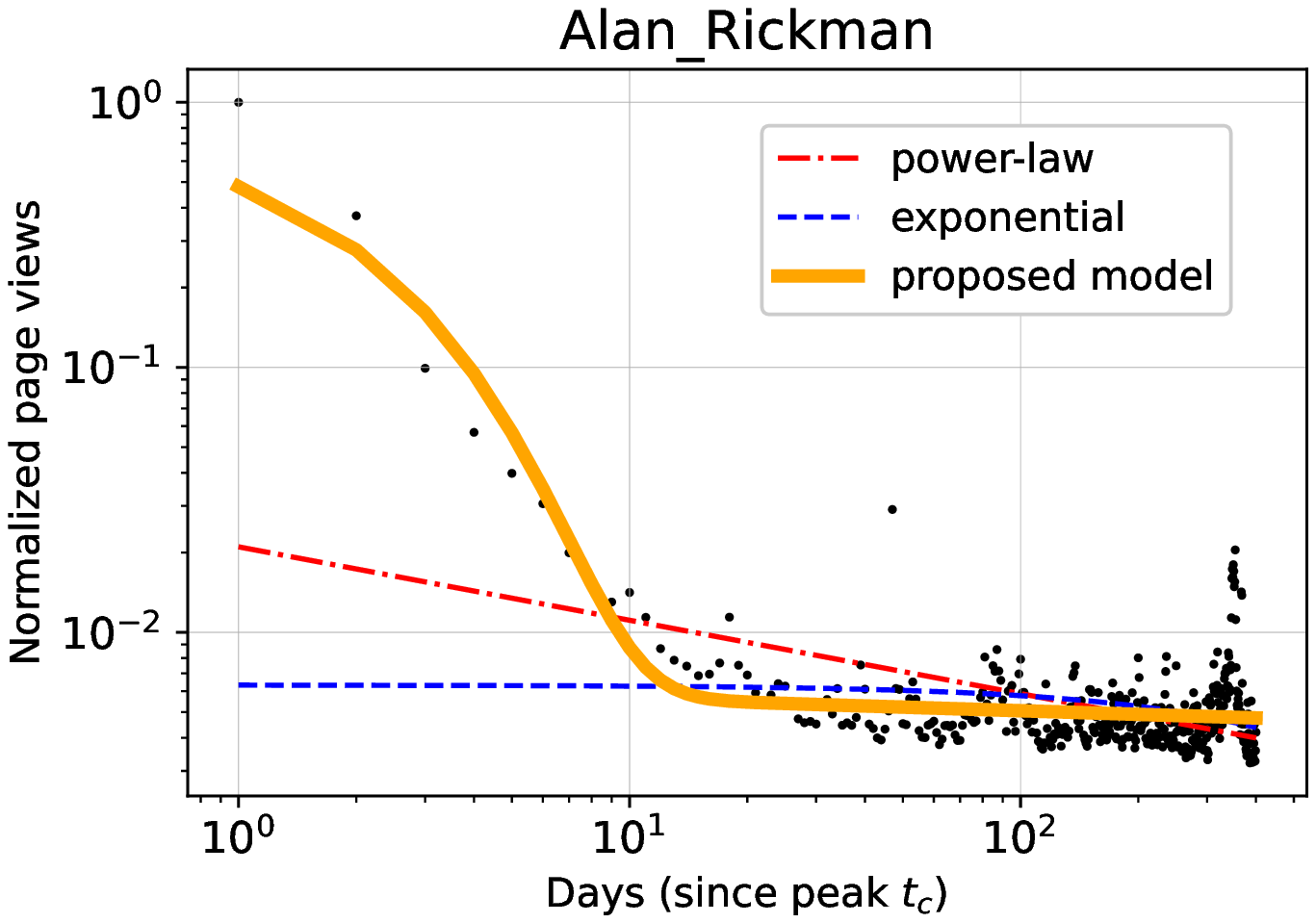}
\caption{Fitting examples of the proposed model.}
\label{Fig3}
\end{figure}

We performed model fitting for each normalized time-series data of page views $S(t)$ with the following 
\red{four nonlinear models that do not assume a regime shift}: bi-exponential $C_{1}\mathrm{e}^{-\alpha t}+C_{2} \mathrm{e}^{-\beta t}$\red{~\cite{candia2019universal}}, stretched exponential $\mathrm{e}^{-(t/\alpha)^{\beta x^{-\gamma}}}$\red{~\cite{kim2021stretched}}, shifted power-law $C_{1} t^{-\alpha}+C_{2}$\red{~\cite{west2021postmortem}}, and the proposed model $C_{1}\mathrm{e}^{-\beta t} + C_{2}t^{-\alpha}$. 
In model fitting, we added a constant value $\epsilon$ to each individual time series, where $\epsilon$ was the minimum nonzero value across all individual time series. Then, we took base-10 logarithms of the empirical data and conducted parameter fitting of each model formula to the data in a log-log space using a nonlinear least-squares method, following the method by West et al.~\cite{west2021postmortem}. 
Figure~\ref{Fig3} shows examples of model fitting. 
Compared to the purely exponential (blue, dashed) and purely power-law (red, dotted) models, our proposed model (orange, solid) can capture both the initial exponential decay and the mid- to long-term power-law decay simultaneously. 

We compared the median of $R^{2}$ and $AIC$ of each model formula for each event category to compare the model performance. Tables \ref{Tab2} and \ref{Tab3} show the results. 

The proposed model showed the best performance for earthquakes, aviation accidents, and terrorist attacks. For deaths of notable persons and mass murder incidents,
the shifted power-law model~\cite{west2021postmortem} performed slightly better, but the differences between its $R^2$ and $AIC$ values and those of our model were small. 
\red{In fact, when we determined each sample individually, our model performed better in more than half of the cases in all categories. 82\%, 59\%, 59\%, 55\%, and 58\% for earthquake, notable death, aviation, mass murder, and terror incidents, respectively.}

\begin{table}[htbp]
\caption{Median of ${R^2}$ for \red{four} decay models.}
      \begin{tabular}{cccccc}
        \hline
        {\tt model}  & {\tt earthquake}  & {\tt death}& {\tt aviation}   & {\tt murder} & {\tt terror} \\ \hline
        bi-exponential \red{~\cite{candia2019universal}}& 0.765 & 0.723 & 0.822 & 0.800 & 0.753\\
        stretched-exponential \red{~\cite{kim2021stretched}} & 0.758 & 0.730 & 0.796 & 0.799 & 0.730\\
        shifted power-law \red{~\cite{west2021postmortem}}& 0.768 & 0.737 & 0.813 & 0.801 & 0.740\\
        proposed model & 0.786 & 0.733 & 0.846 & 0.798 & 0.757\\
        \hline
    \end{tabular}
    \label{Tab2}  
\end{table}

\begin{table}[htbp]
\caption{Median of $AIC$ for \red{four} decay models.}
      \begin{tabular}{cccccc}
        \hline
        {\tt model}  & {\tt earthquake}  & {\tt death}& {\tt aviation}   & {\tt murder} & {\tt terror} \\ \hline
        bi-exponential \red{~\cite{candia2019universal}} & -277.0 & -229.1 & -423.1 & -269.7 & -244.5\\
        stretched-exponential \red{~\cite{kim2021stretched}} & -284.4 & -244.5 & -408.7 & -279.3 & -227.7\\
        shifted power-law \red{~\cite{west2021postmortem}} & -294.0 & -250.1 & -412.5 & -281.3 & -240.6\\
        proposed model& -311.1 & -245.4 & -436.9 & -280.9 & -262.1\\
        \hline
    \end{tabular}
    \label{Tab3}  
\end{table}

\subsection{Decay parameters}
\begin{figure}
\begin{minipage}[t]{0.45\linewidth}
    \centering
    \includegraphics[scale=0.35]{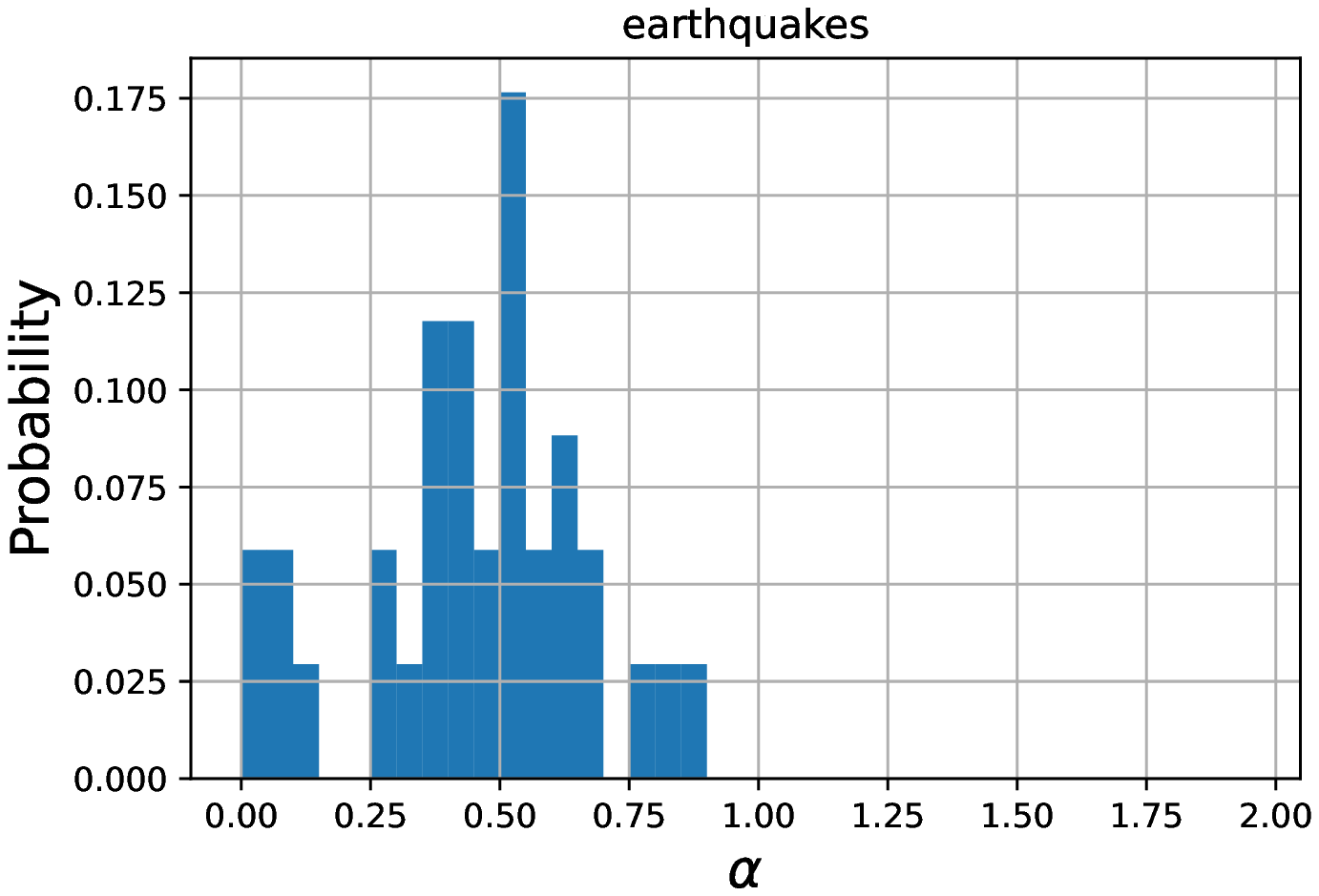}
  \end{minipage}
  \begin{minipage}[t]{0.45\linewidth}
    \centering
    \includegraphics[scale=0.35]{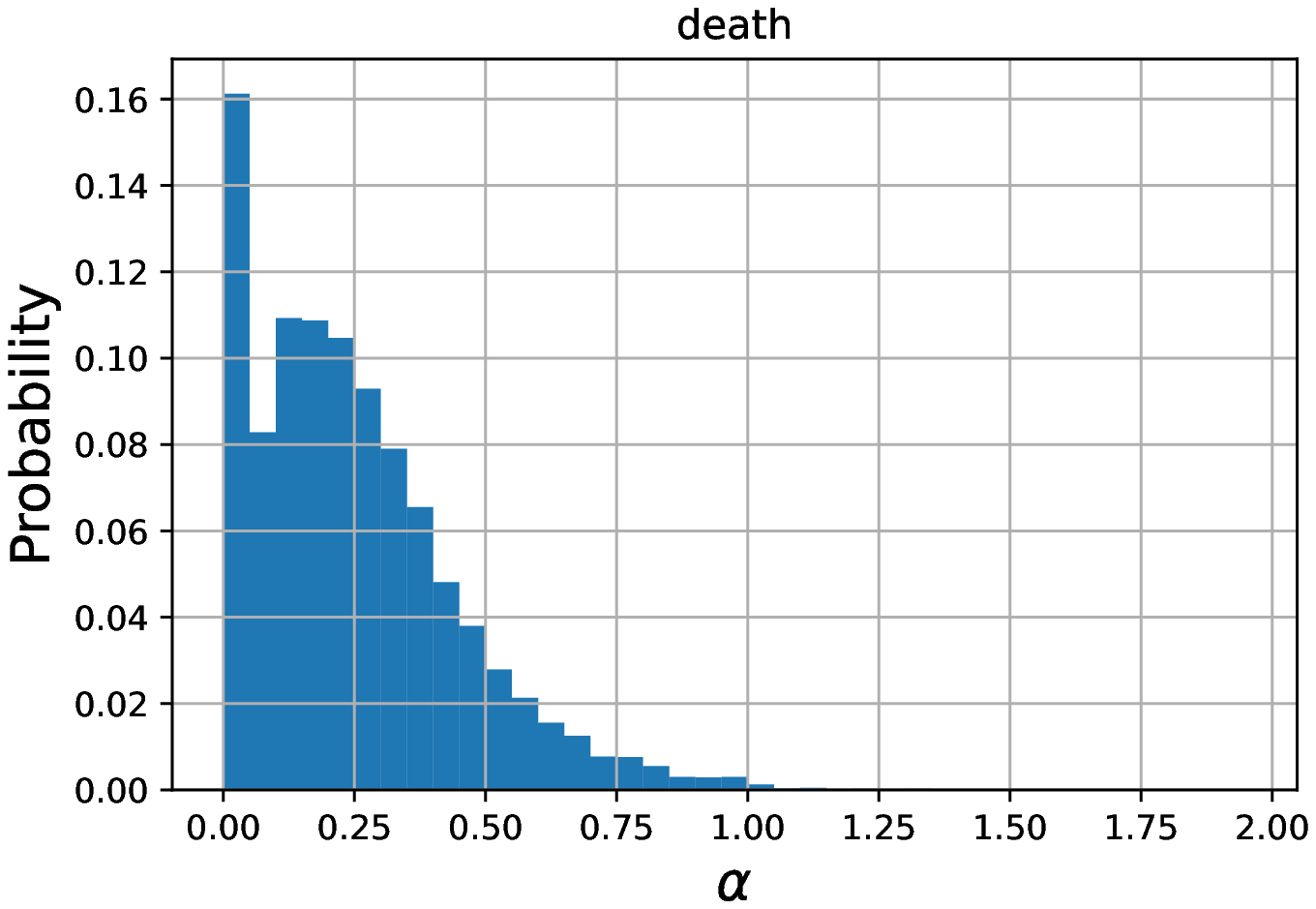}
  \end{minipage}
  \begin{minipage}[t]{0.45\linewidth}
    \centering
    \includegraphics[scale=0.35]{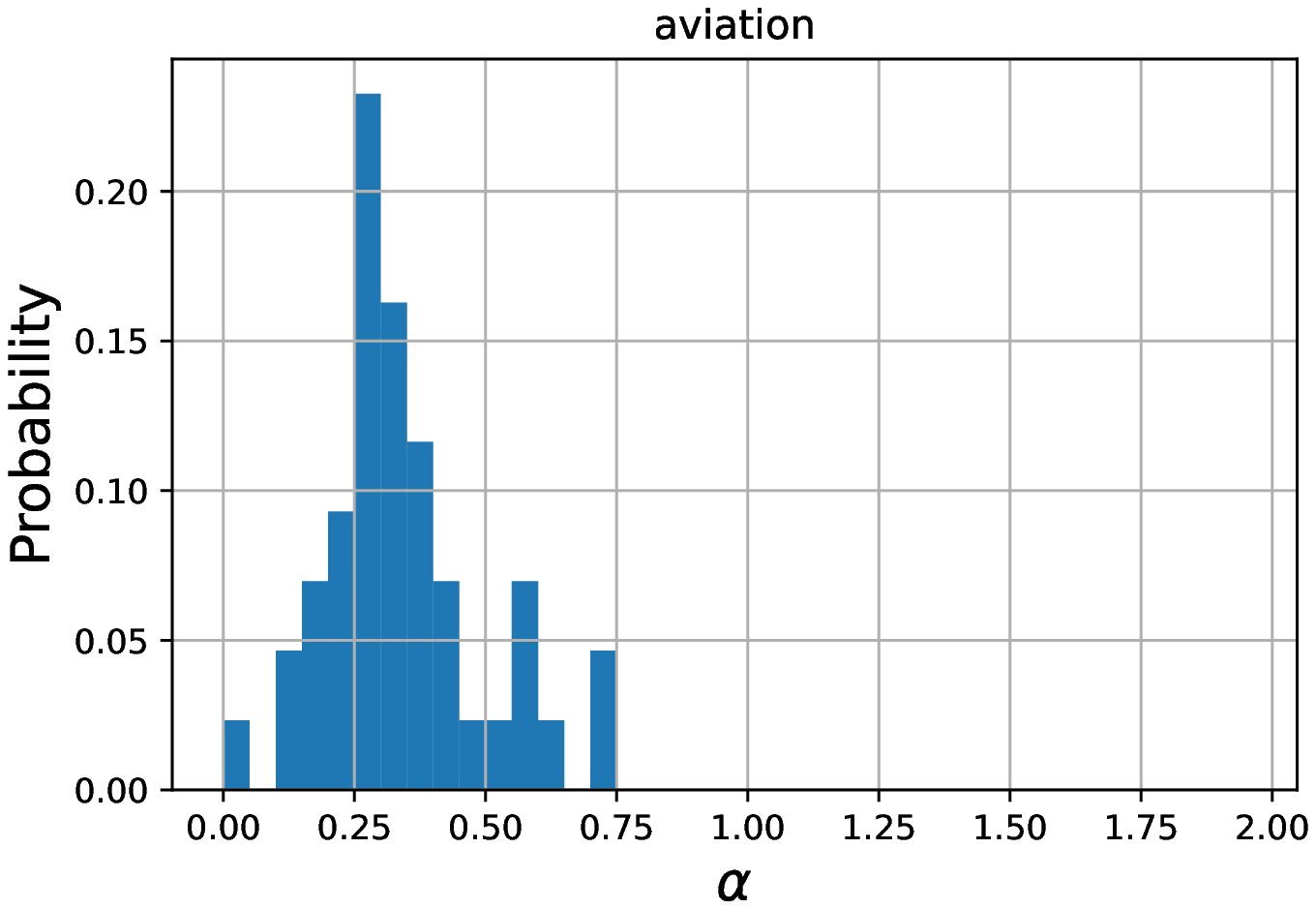}
  \end{minipage}
  \begin{minipage}[t]{0.45\linewidth}
    \centering
    \includegraphics[scale=0.35]{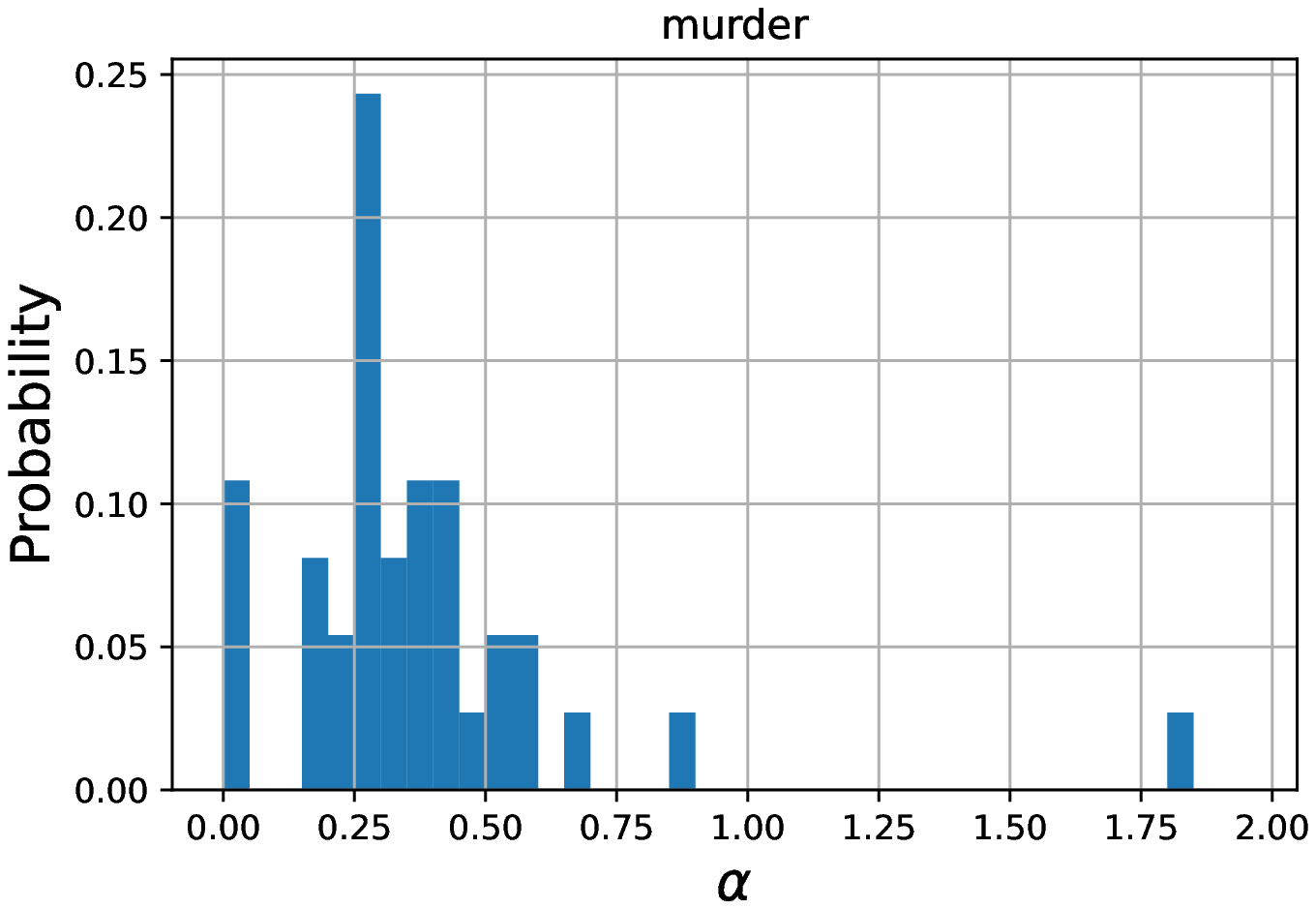}
  \end{minipage}
  \begin{minipage}[t]{0.45\linewidth}
    \centering
    \includegraphics[scale=0.35]{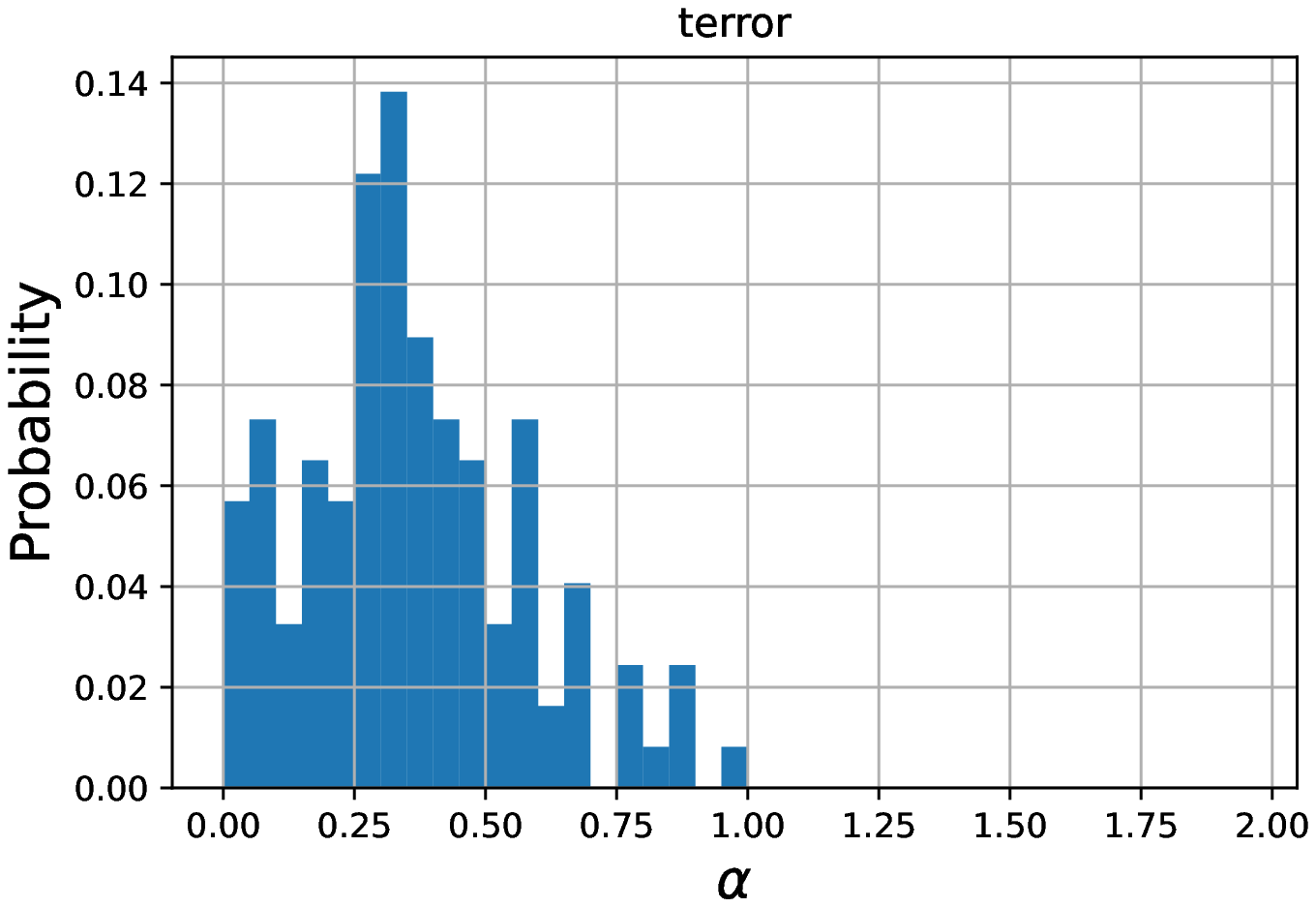}
  \end{minipage}
\caption{Probability density distributions of parameter values of $\alpha$ obtained using the proposed model.}
\label{Fig5a}
\end{figure}

\begin{figure}
\begin{minipage}[t]{0.45\linewidth}
    \centering
    \includegraphics[scale=0.35]{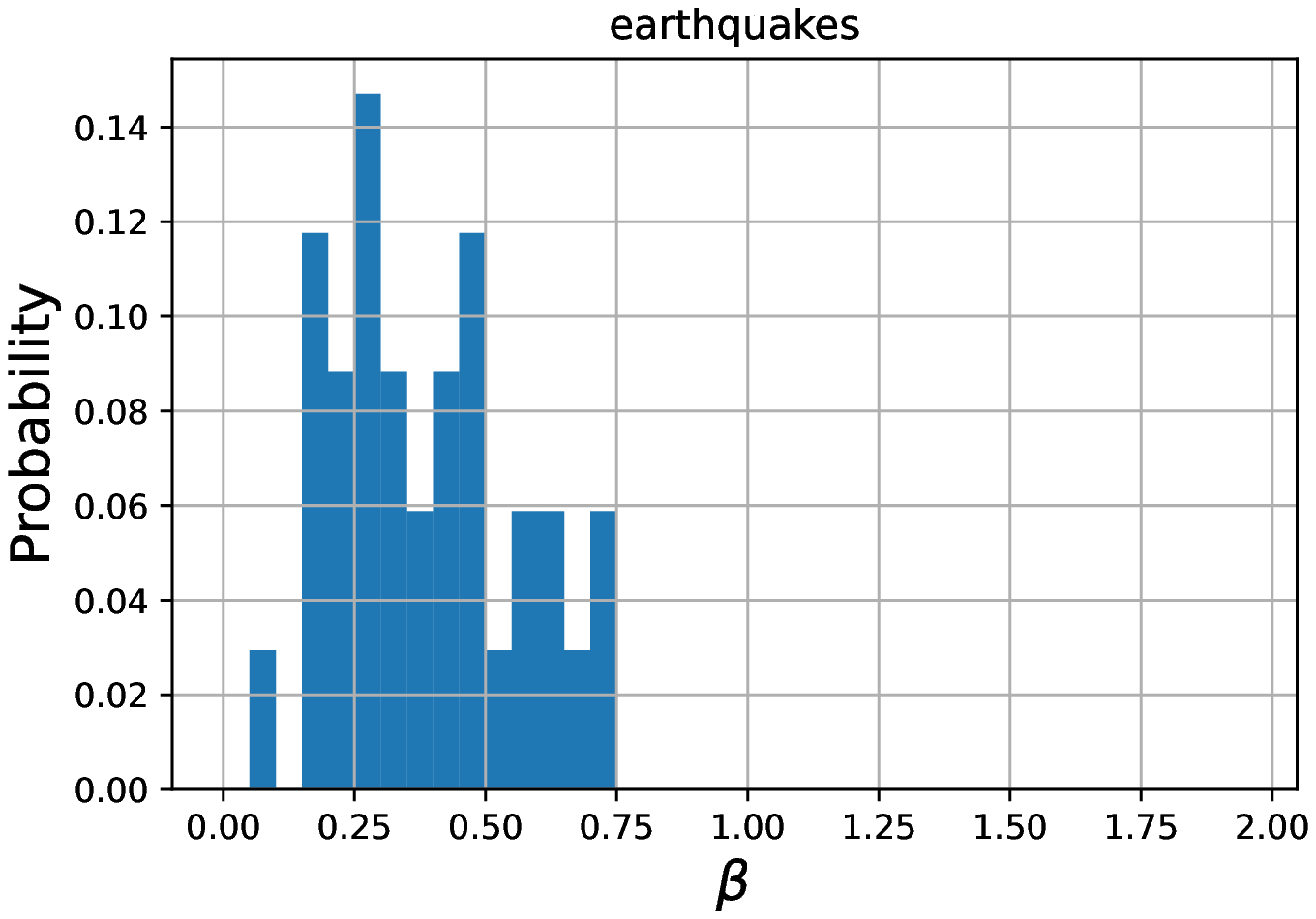}
  \end{minipage}
  \begin{minipage}[t]{0.45\linewidth}
    \centering
    \includegraphics[scale=0.35]{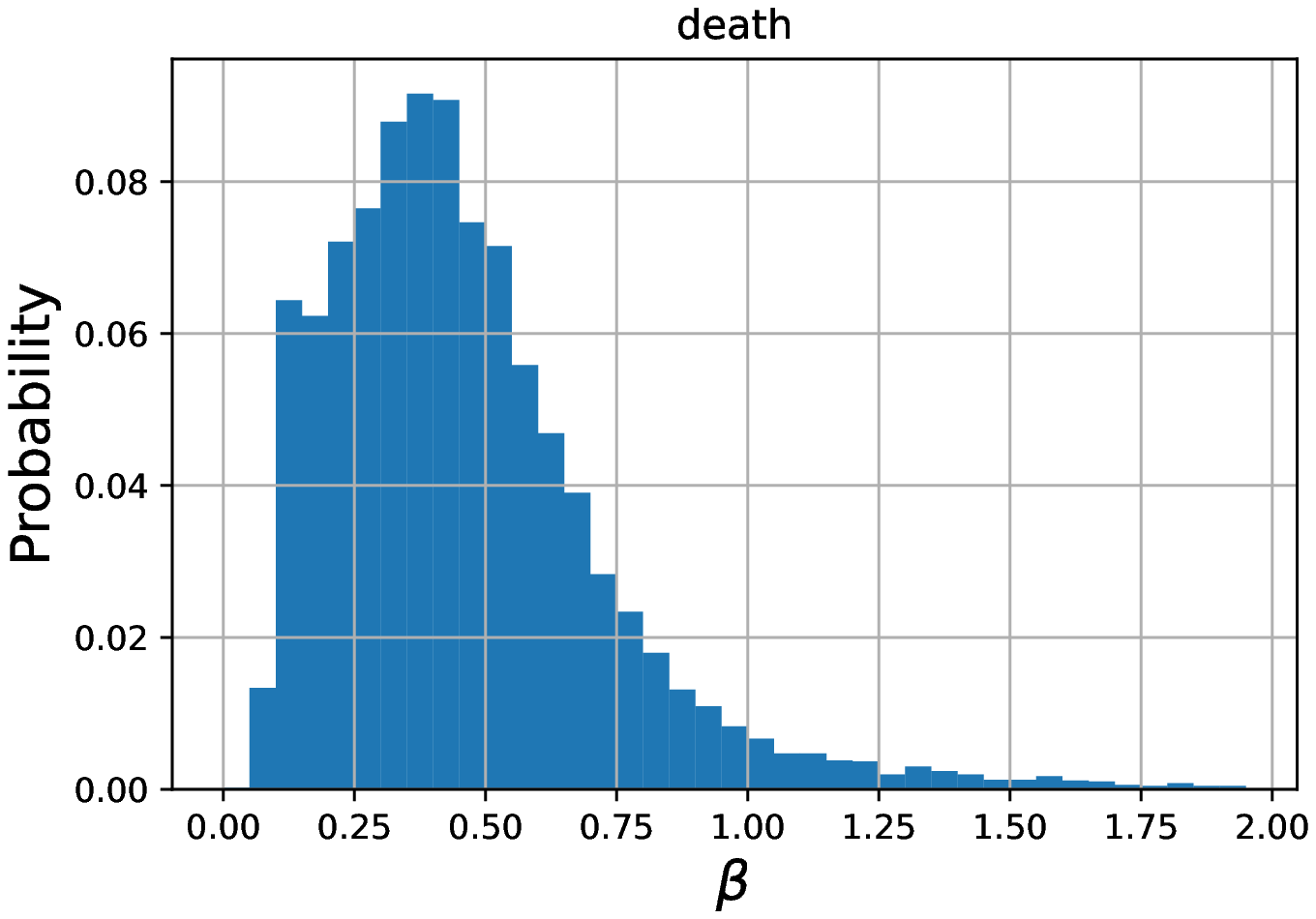}
  \end{minipage}
  \begin{minipage}[t]{0.45\linewidth}
    \centering
    \includegraphics[scale=0.35]{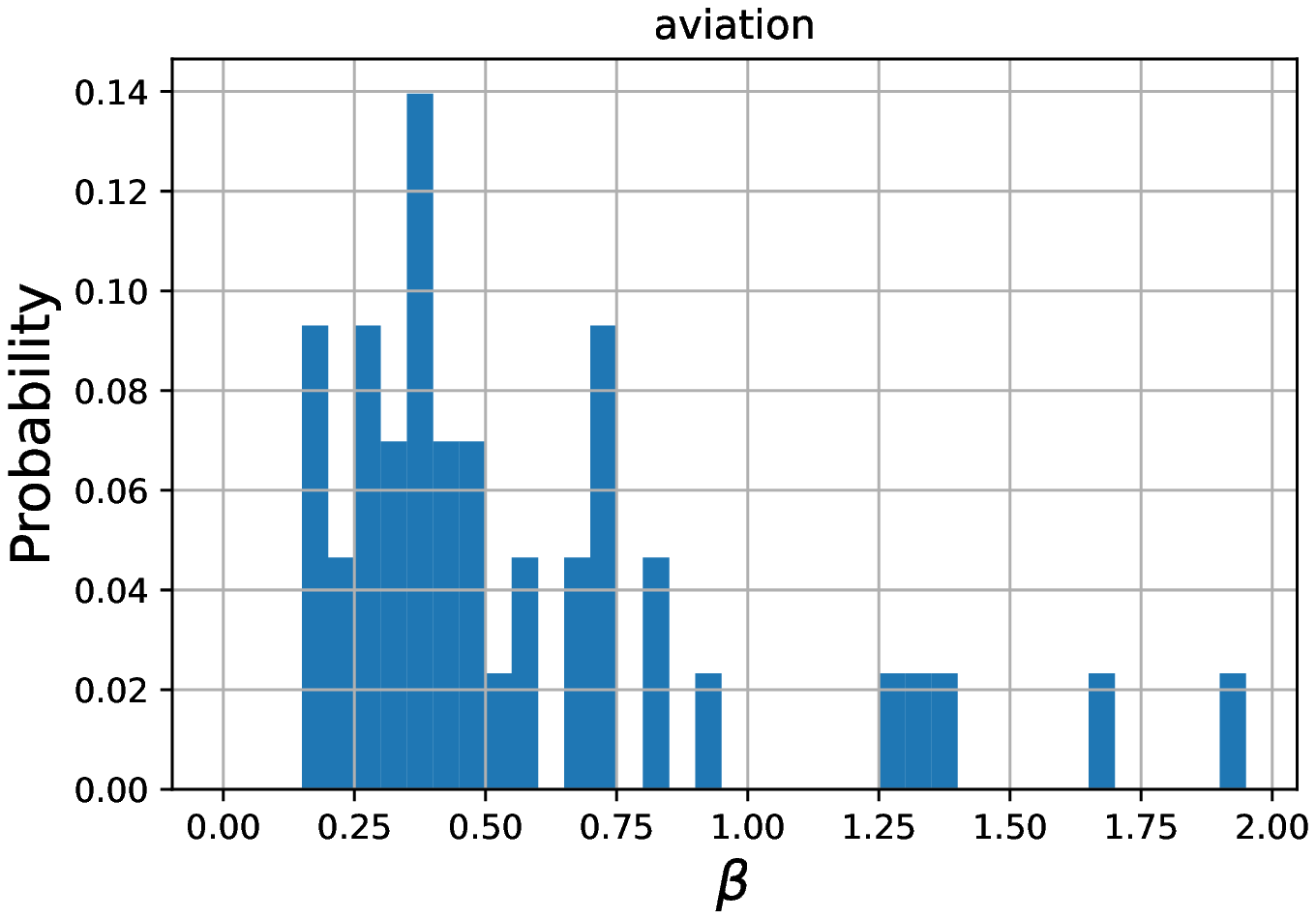}
  \end{minipage}
  \begin{minipage}[t]{0.45\linewidth}
    \centering
    \includegraphics[scale=0.35]{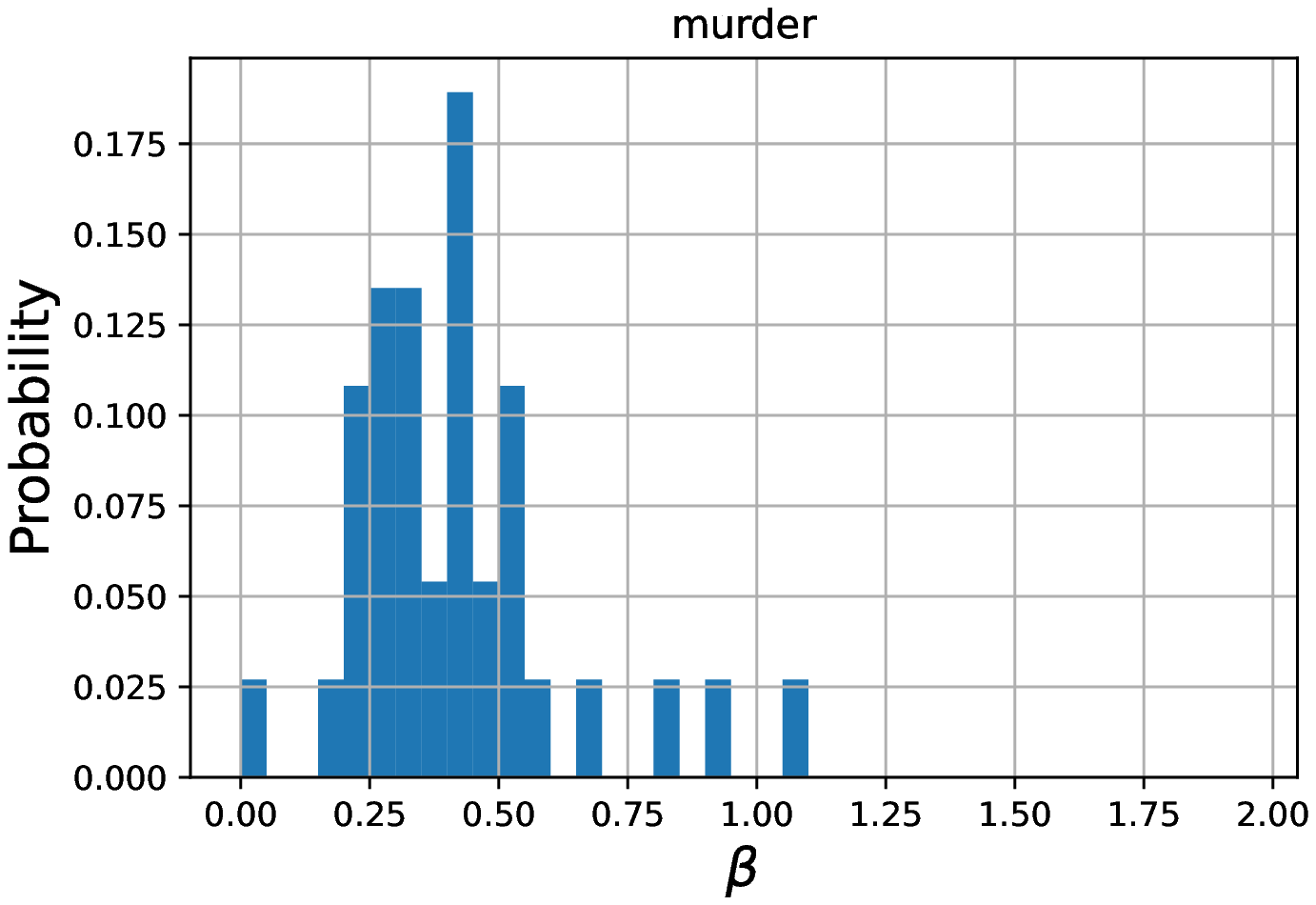}
  \end{minipage}
  \begin{minipage}[t]{0.45\linewidth}
    \centering
    \includegraphics[scale=0.35]{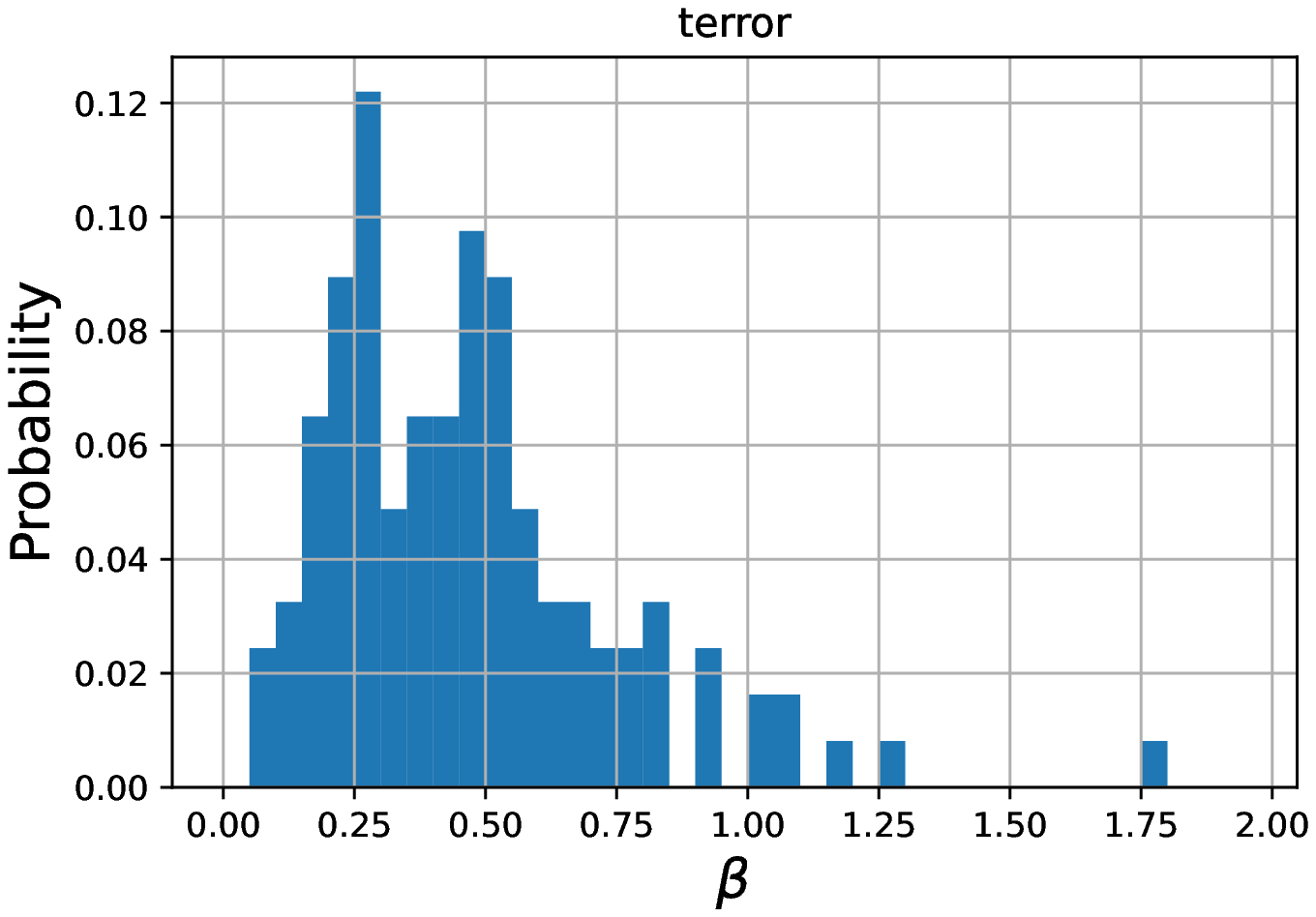}
  \end{minipage}
\caption{Probability density distributions of parameter values of $\beta$ obtained using the proposed model.}
\label{Fig5b}
\end{figure}

The initial fast exponential decay is characterized by $\beta$ and the late slow power-law decay by $\alpha$. Figure\red{s}~\ref{Fig5a} \red{and~\ref{Fig5b}} show probability density distributions of the parameter values of $\beta$ and $\alpha$. Note that 26 (0.3\%) outliers ($\beta > 2$) are not shown in the distribution \red{of death of notable persons ($N=8,684$)}. 
These distributions show a clear unimodal distribution with a distinct characteristic value for each parameter \red{whose medians are shown in Table~\ref{Tab4}.}
These results suggest that there is a \red{common} pattern of collective memory decay, first in the fast exponential decay immediately after the event with the exponent $\beta$ around 0.4, followed by the slow power-law decay with the exponent $\alpha$ around 0.3. 

An interesting observation is that the value of $\alpha$ may be loosely related to the lasting societal impact of the events. Earthquakes tend to cause a massive damage to society and the characteristic value of $\alpha$ for this category was large (0.48), implying that there was meaningful long-term collective memory decay going on for a long period of time. Meanwhile, deaths of notable persons would have minimal impact on society and its characteristic value of $\alpha$ was small (0.22), implying that the long-term behavior was closer to a flat line ($\alpha=0$) and more likely dominated by constant random page views. Events in other categories would have societal impacts at intermediate levels, which may be reflected on their intermediate characteristic $\alpha$ values as well. This observation remains largely speculative and would need further systematic investigation.

\clearpage

\clearpage

\begin{table}[htbp]
\caption{\red{Median of $\beta$ and $\alpha$ for five categories for the proposed model}.}
      \begin{tabular}{cccccc}
        \hline
          & {\tt earthquake}  & {\tt death}& {\tt aviation}   & {\tt murder} & {\tt terror} \\ \hline
        $\beta$ & 0.39 & 0.42 & 0.45 & 0.41 & 0.43\\
        $\alpha$ & 0.48 & 0.22 & 0.31 & 0.30 & 0.33\\
        \hline
    \end{tabular}
    \label{Tab4}  
\end{table}

\subsection{Switching point of collective memory decay dynamics}
\begin{figure}[htbp]
    \includegraphics[scale=0.5]{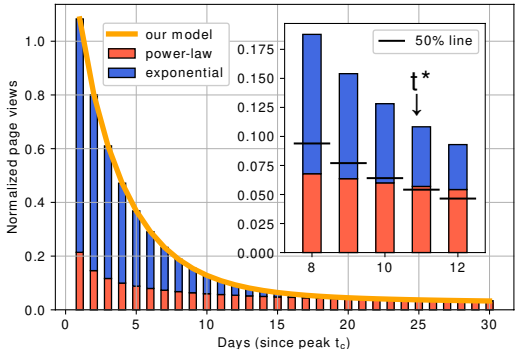}
    \caption{Overview of the detection of the collective memory switching point $t^*$.}
\label{Fig6}
\end{figure}
Our proposed model allows for detection of the ``switching point'' of collective memory decay where the dominant component in the model formula $S_{i}(t) =C_{1}\mathrm{e}^{-\beta t} + C_{2}t^{-\alpha}$ switches from exponential to power-law. 
Such a switching point $t^{*}$ is defined as the first time point at which $C_{2}t^{-\alpha} > C_{1}\mathrm{e}^{-\beta t}$ in the fitted model (Fig.~\ref{Fig6}).

\begin{figure}
  \begin{minipage}[t]{0.45\linewidth}
    \centering
    \includegraphics[scale=0.35]{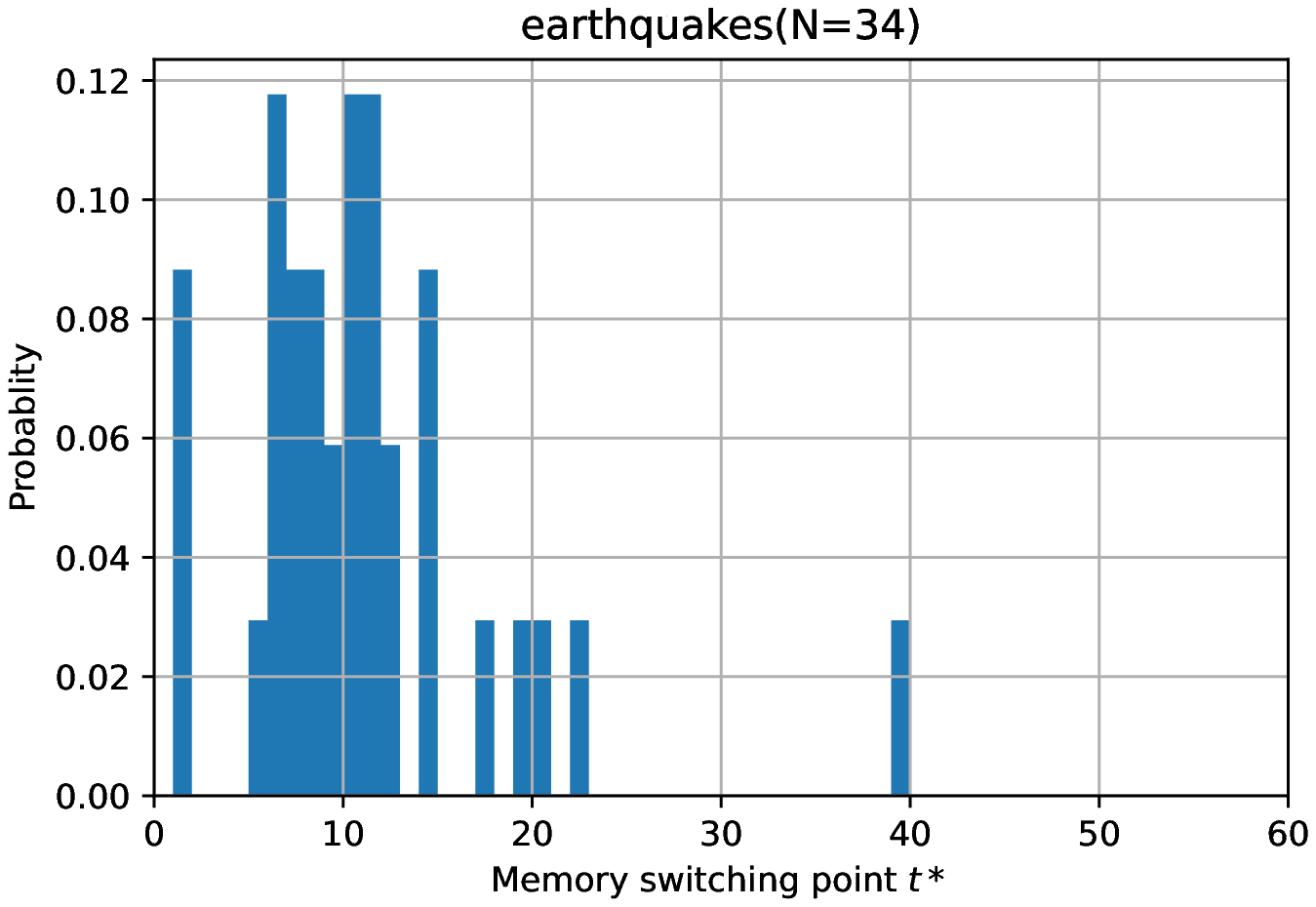}
  \end{minipage}
  \begin{minipage}[t]{0.45\linewidth}
    \centering
    \includegraphics[scale=0.35]{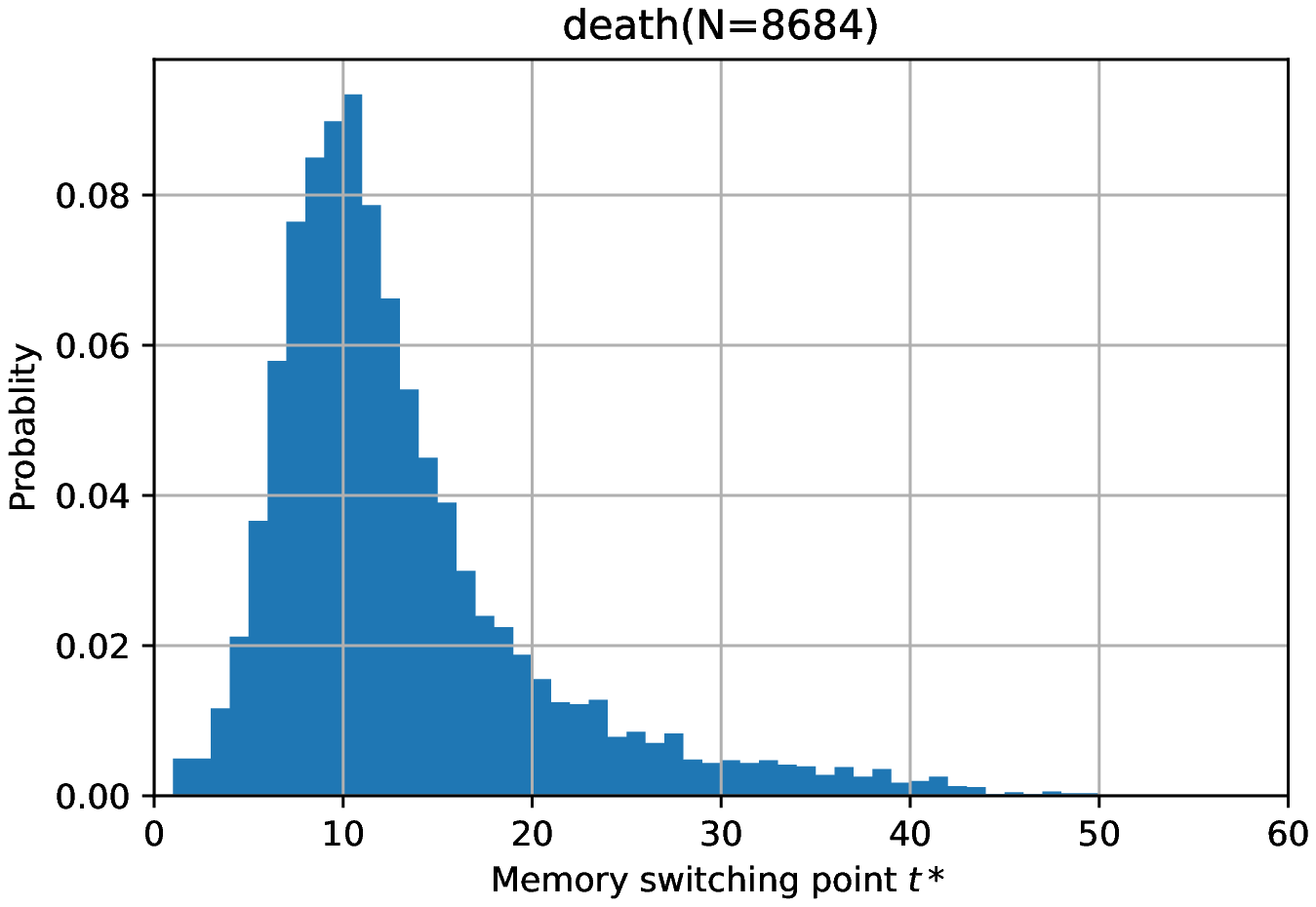}
  \end{minipage}
  \begin{minipage}[t]{0.45\linewidth}
    \centering
    \includegraphics[scale=0.35]{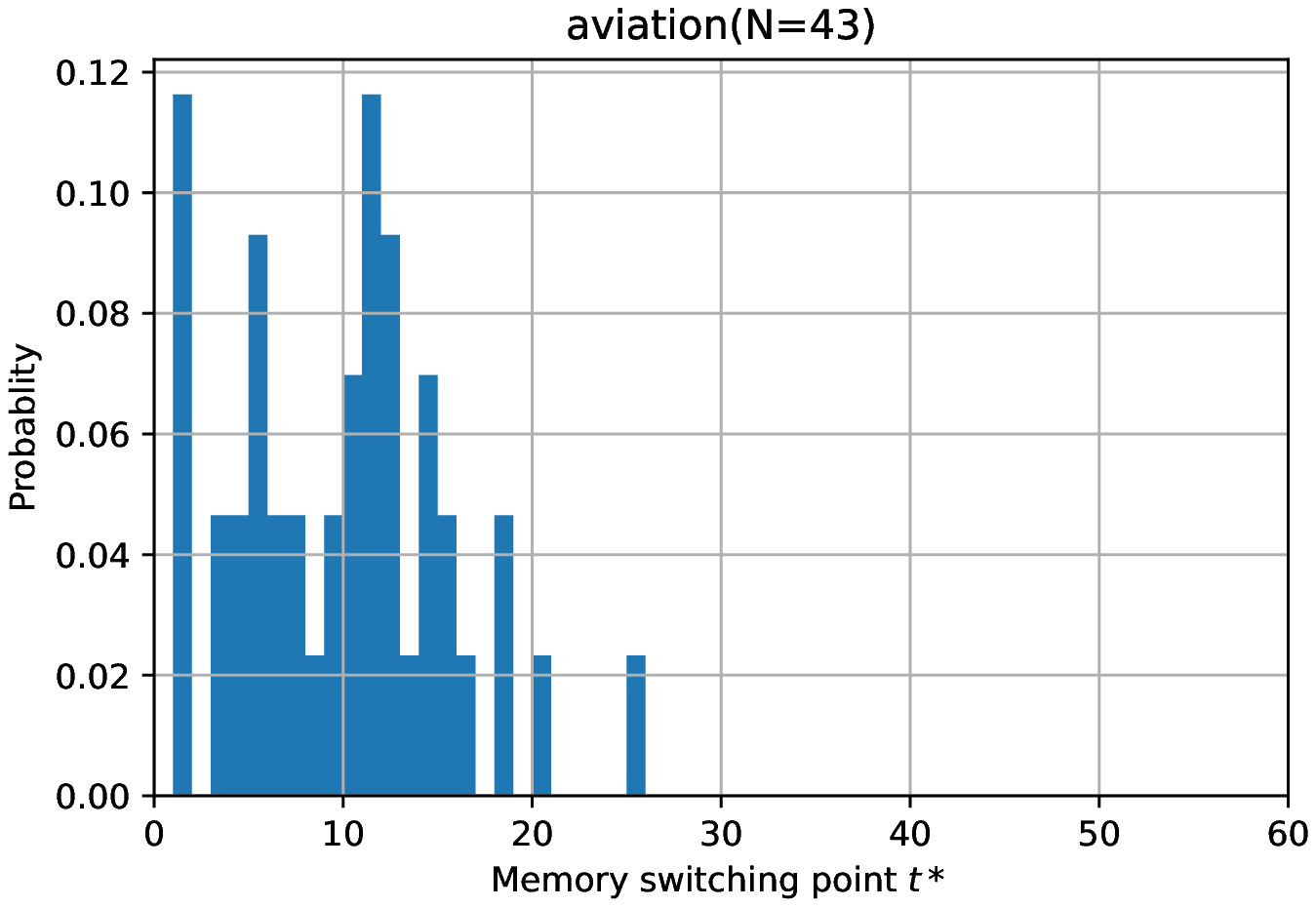}
  \end{minipage}
  \begin{minipage}[t]{0.45\linewidth}
    \centering
    \includegraphics[scale=0.35]{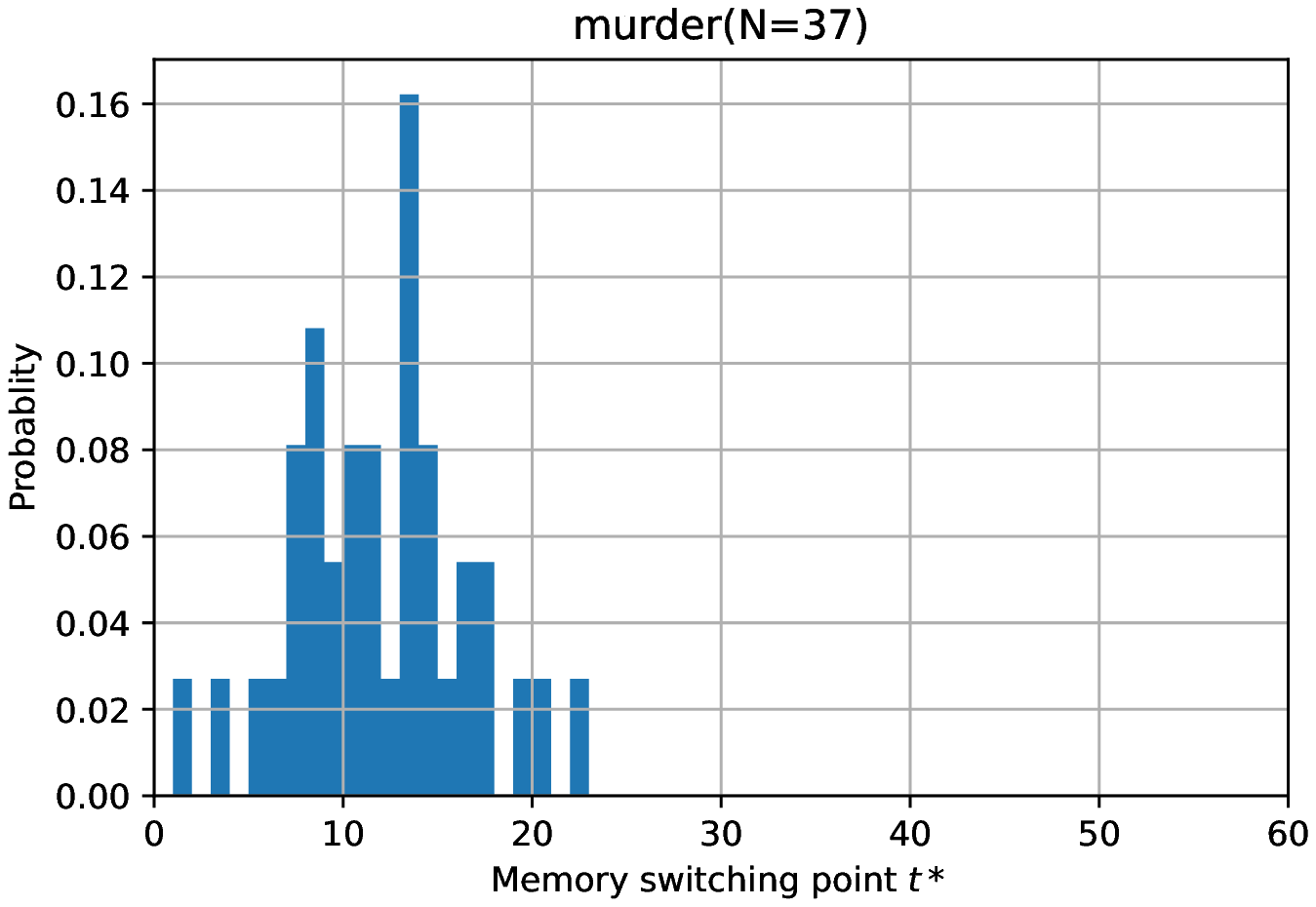}
  \end{minipage}
  \begin{minipage}[t]{0.45\linewidth}
    \centering
    \includegraphics[scale=0.35]{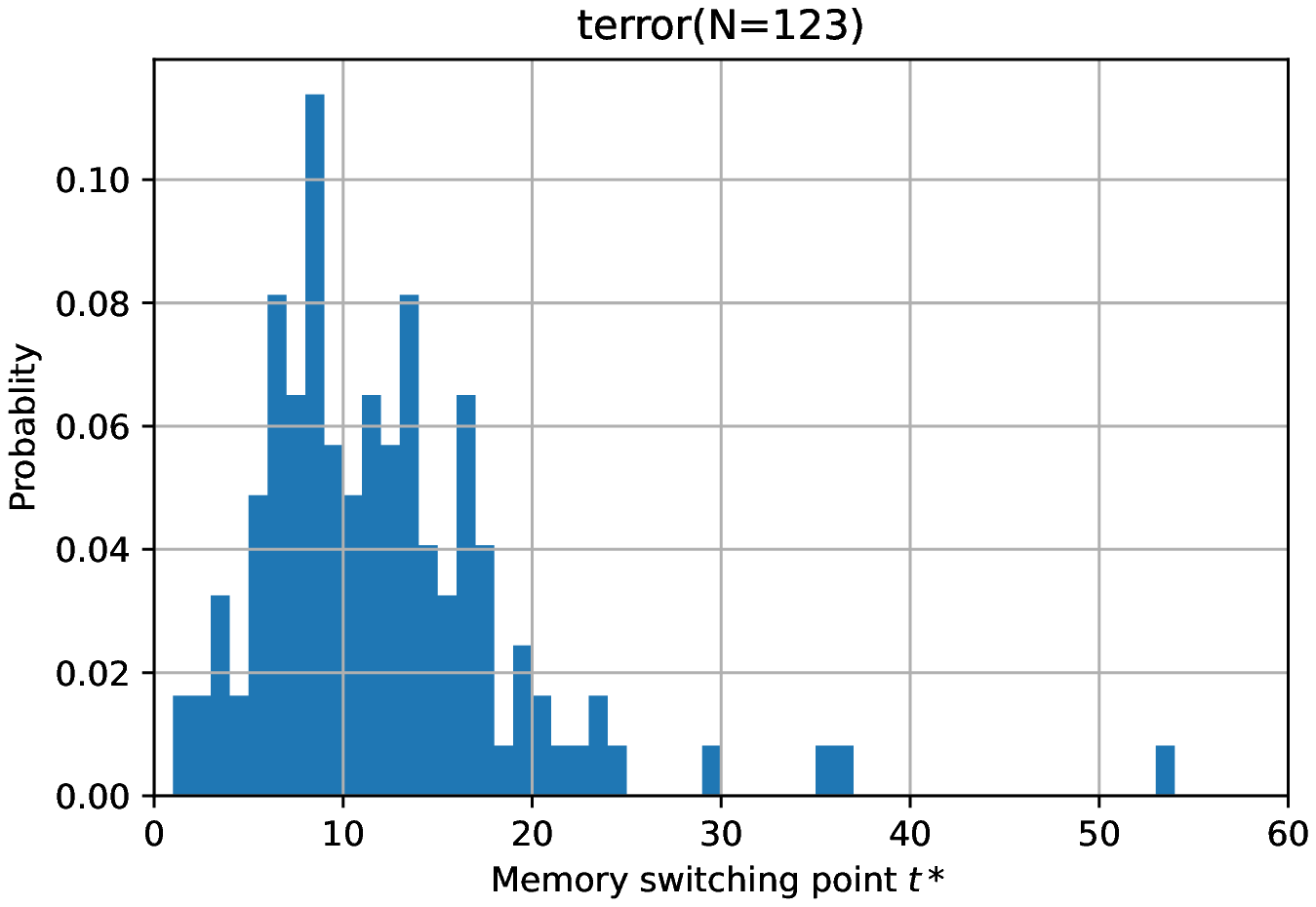}
  \end{minipage}
\caption{Probability density distributions of switching points $t^*$ detected for five categories.}
\label{Fig7}
\end{figure}

Figure~\ref{Fig7} shows the probability density distributions of the switching points detected for five categories. The median values for all categories were quite similar (earthquake: $t^{*}=10$; deaths of notable persons: $t^{*}=11$; aviation accidents: $t^{*}=10$; mass murder incidents: $t^{*}=11$; and terrorist attacks: $t^{*}=11$), indicating a  \red{common} pattern of the shift of collective memory decay dynamics at about the same timing (around 10 to 11 days after the peak), regardless of the event category.

\clearpage

\section{Discussions}
In this study, we collected daily English Wikipedia page view counts for five event categories and modeled their decay processes using a new two-phase model that combined initial exponential decay and mid- to long-term power-law decay in a single mathematical formula. To the limit of our knowledge, this study was the first attempt to develop a universal model of collective memory decay applicable to multiple event categories at daily time scales.
We found that our proposed model showed consistently high accuracy across multiple event categories, and closely matching the best performance in \red{the previously proposed decay models}. 

Our model also allowed for the detection of a ``switching point'' in collective memory decay at which the dominant decay dynamics switches from exponential to power-law. We found that the decay phase switches about 10 to 11 days after the peak, irrespective of the event category. 
\red{This number is similar to what was reported in Garc\'{i}a-Gavilanes et al.~\cite{garcia2016dynamics} that the first break point of the segmentation was 3-10 days for both English and Spanish Wikipedia page views of aviation accidents.}
This is a unique, non-trivial finding because it indicates a universal property of our society's ``collective attention span'' which shows immediate attention period of the news.

\red{There are some limitations in our study. Firstly, we only validated our model by the English Wikipedia page views. 
Therefore one still needs to be careful in considering generality of the obtained results by using other data like Twitter mentions.
The assumptions of the model should also be noted. Here we focused on aggregated behavior of the user population, and we did not consider each individual user's behavioral changes. }

Future directions of research include consideration of more detailed information about specific events and modeling their influences on collective memory decay, such as more detailed event types, the popularity of the event, and the size of societal impacts the event created. Such systematic analysis will help understand the nature of collective memory in greater depth, possibly revealing the quantitative relationship between the event's impact and the value of $\alpha$ as indicated above. 
\red{Also, we found the spontaneous increases in collective memory decay around 365 days which could be attributed to year-to-year recall. We recognize that investigating such spontaneous increases is another interesting future direction.}

\section*{Competing interests}
  The authors declare that they have no competing interests.

\section*{Availability of data}
    The datasets used and analyzed during the current study are available through the Wikimedia REST API.

 \bibliographystyle{elsarticle-num} 
 \bibliography{cas-refs}

\end{document}